\documentclass[preprint,aps,amsmath,amssymb,nofootinbib,12pt]{revtex4}

\usepackage{epsfig}
\usepackage{slashed}
\usepackage{graphicx}
\usepackage{multirow,color}
\usepackage{amsmath}
\usepackage{float}
\usepackage{diagbox}
\usepackage{CJK}
\usepackage{color}
\usepackage{xcolor}
\usepackage{times}
\usepackage{subfigure}
\usepackage{bm}
\usepackage{braket}
\usepackage{booktabs}
\usepackage{array}
\usepackage[mathscr]{euscript}
\usepackage{caption}
\usepackage{makecell}
\usepackage{epstopdf}
\usepackage{hyperref}
\usepackage{amsmath,amssymb,amsthm,amsxtra,overpic,bbm,bm,epsfig,ulem,color,multirow}

\makeatletter

\newcommand{\Rmnum}[1]{\expandafter\@slowromancap\romannumeral #1@}
\makeatother
\graphicspath{{fig/}}
\begin{document}
\title{Searching for long-lived axion-like particles via displaced vertices at the HL-LHC}
\author{Chong-Xing Yue$^{1,2}$}
\thanks{cxyue@lnnu.edu.cn}
\author{Xin-Yang Li$^{1,2}$}
\thanks{lxy91108@163.com~(Corresponding author)}
\author{Shuo Yang$^{1,2}$}
\thanks{shuoyang@lnnu.edu.cn}
\author{Mei-Shu-Yu Wang$^{1,2}$}
\thanks{wmsy0108@163.com}

\affiliation{
$^1$Department of Physics, Liaoning Normal University, Dalian 116029, China\\
$^2$Center for Theoretical and Experimental High Energy Physics, Liaoning Normal University, Dalian 116029, China
}

\begin{abstract}
Axion-like particles~(ALPs) are well-motivated extensions of the standard model~(SM) that appear in numerous new physics scenarios. In this paper, we concentrate on searches for long-lived ALPs predicted by the photophobic scenario at the HL-LHC with the center-of-mass energy $\sqrt{s}=14$ TeV and the integrated luminosity $\mathcal{L}=$ $3$ ab$^{-1}$. We consider the process $pp \to \gamma a$ with the ALP $a$ decaying into a pair of displaced charged leptons and perform a detailed analysis of two types of signals: $\pi^+ \pi^- \gamma \slashed{E}_T $ and $\ell^+ \ell^- \gamma \slashed{E}_T $. For the $\pi^+ \pi^- \gamma \slashed{E}_T $ signal, we find that the prospective sensitivities of the HL-LHC can reach $g_{aWW}\in [8.72 \times 10^{-3}, 6.42 \times 10^{-2}]$ TeV$^{-1}$ for the ALP mass $m_a \in [4, 10]$ GeV. While for the $\ell^+ \ell^- \gamma \slashed{E}_T $ signal, the HL-LHC can probe a broader parameter space, with the sensitivities covering $m_a \in [4, 10]$ GeV and $g_{aWW} \in [4.17 \times 10^{-3}, 2.00 \times 10^{-1}]$ TeV$^{-1}$. These long-lived searches complement some previous prompt decay studies from LEP and LHC experiments, extending the parameter space explored by the LHCb collaboration. Our results show that the HL-LHC has significant potential to probe long-lived ALPs via their displaced vertex signatures.
\end{abstract}
\maketitle

%%%%%%%%%%%%%%%%%%%%%%%%%%%%%%%%%%%%%%%%%%%%%%%%%%
%%%%%%%%%%%%%%%%%%%%%%%%%%%%%%%%%%%%%%%%%%%%%%%%%%
\section{Introduction}
%%%%%%%%%%%%%%%%%%%%%%%%%%%%%%%%%%%%%%%%%%%%%%%%%%
%%%%%%%%%%%%%%%%%%%%%%%%%%%%%%%%%%%%%%%%%%%%%%%%%

Axion-like particles~(ALPs) naturally emerge in extensions of the standard model~(SM) of particle physics as CP-odd pseudo Nambu-Goldstone bosons associated with the spontaneous breaking of global symmetries, which are singlets under the SM gauge groups. ALPs can serve
as compelling candidates for dark matter~\cite{Preskill:1982cy,Abbott:1982af,Dine:1982ah} or inflaton~\cite{Freese:1990rb}, and provide solutions to the gauge hierarchy problem~\cite{Graham:2015cka} as well as the neutrino mass problem~\cite{Dias:2014osa,Chen:2012baa,Salvio:2015cja}. In addition, ALPs have been proposed to account for several observed phenomena, such as the matter-antimatter asymmetry~\cite{Jeong:2018jqe,Co:2019wyp} and the anomalous magnetic dipole moment of muon~\cite{Bauer:2019gfk,Marciano:2016yhf}. In general, ALPs can couple to all SM particles with model-dependent couplings, which lead to a multitude of possible signatures at various experiments. Consequently, ALPs can be searched through astrophysical and cosmological observations as well as laboratory experiments.

ALP generalizes the concept of the QCD axion~\cite{Wilczek:1977pj,Weinberg:1977ma,Peccei:1977hh,Peccei:1977ur}, but without being constrained by the strict mass-coupling relation, allowing its mass $m_a$ and coupling strength suppressed by the global symmetry breaking scale $f_a$ are independent parameters.  This flexibility enables ALP to interact with various particles and span a wide mass range, which are subject to stringent constraints from various experiments. Cosmological and astrophysical observations~(see e.g.~\cite{Marsh:2015xka,Planck:2015fie,Caputo:2024oqc}) provide stringent bounds on very light ALPs with masses below the MeV scale. For heavier ALPs with masses from several MeVs to hundreds of GeV, high-energy colliders offer sensitive probes, where the detectability of ALP signals strongly depends on their lifetimes.
Specifically, heavier ALPs or those with stronger couplings tend to be short-lived, decaying promptly into the SM particles within the detector and producing resonant signatures. Many studies have explored such prompt decays to probe ALP couplings to the SM fermions and gauge bosons at hadron and lepton colliders, for example see Refs.~\cite{Bauer:2017ris,Bauer:2017nlg,CMS:2017dmg,Inan:2020aal,Inan:2020kif,Zhang:2021sio,Yue:2021iiu,
Yue:2023mew,Yue:2023mjm,Han:2022mzp,RebelloTeles:2023uig}. In contrast, ALPs with weaker couplings or lighter masses can be sufficiently long-lived and escape the detector before decaying, leading to missing energy signature, which provides an alternative avenue for probing the ALP interactions, for instance see Refs.~\cite{Mimasu:2014nea,Brivio:2017ije,Chenarani:2025uay,Bao:2025tqs}.

Beyond the well-studied cases of prompt decays and missing energy signatures, an intermediate region arises when the ALP lifetime appropriately matches the detector size. In this case, ALPs can travel a macroscopic distance from their production point before decaying, leading to displaced vertex~(DV) signatures within the detector, which offer a unique signal to probe the ALPs with intermediate lifetimes, complementing previous searches focused on prompt decays and missing energy signatures. The Large Hadron Collider~(LHC), with high center-of-mass energy, excellent spatial resolution, and precise tracking capabilities, is well suited to probe long-lived ALPs through DV signatures. Refs.~\cite{ATLAS:2024qoo,Ghebretinsaea:2022djg,Cheung:2024qve} present dedicated searches for long-lived ALPs decaying into hadrons with significantly displaced vertices. In addition to hadronic decays, displaced decays into charged leptons have also emerged as a promising channel for probing long-lived ALPs~\cite{Rygaard:2023dlx,Shan:2024pcc}.

The photophobic ALP scenario~\cite{Craig:2018kne} predicts distinctive and novel collider signatures that differ from conventional ALP phenomenology. These ALPs are characterized by strongly suppressed couplings to diphotons, with their primary interactions occurring with other electroweak gauge bosons~($WW,ZZ$ and $Z\gamma$). Previous searches~\cite{Aiko:2024xiv,Ding:2024djo,Mao:2024kgx} have predominantly focused on heavy photophobic ALPs, lighter ones also exhibit rich phenomenology. In the low-mass region, ALP decays to massive gauge bosons are kinematically forbidden, while the dominant decay channels being light fermions and photons. Although ALPs do not couple directly to fermions or photons at tree level, the effective ALP-fermion and ALP-photon couplings can be induced at loop level involving electroweak gauge bosons. These loop-induced couplings, being suppressed, can significantly extend the lifetimes of ALPs, allowing them to decay into light particles with significant displacement from their production point, leading to unique collider signatures. Moreover, these couplings can be expressed in terms of ALP-$W$ boson couplings. Although numerous displaced lepton and hadron searches have been performed at high-energy colliders, their implications for the parameter space of the photophobic ALP scenario have not been systematically explored. Motivated by these, we investigate the potential displaced vertex signatures of long-lived ALPs within the photophobic ALP framework and explore regions of parameter space that remain unconstrained by current experiments.

The High-Luminosity LHC~(HL-LHC)~\cite{Apollinari:2017lan,ZurbanoFernandez:2020cco} has better potential to detect the DV signatures of long-lived particles, benefiting from higher integrated luminosity and improved detector performance than the LHC. Its general-purpose detectors, such as ATLAS and CMS, are equipped with key sub-detectors, including the inner tracking detector~(ID), electromagnetic calorimeter~(ECAL), hadronic calorimeter~(HCAL), and the muon system~(MS), which enable accurate tracking and particle identification. Among these, the ID plays a crucial role in effectively detecting DV signatures, owing to its high-precision tracking and ability to accurately measure the transverse impact parameters of charged decay products. ALPs with sufficiently long lifetimes may fly out of the ATLAS or CMS detectors and reach the far detectors, which can also be detected, for example see Refs.~\cite{Feng:2018pew,Kling:2022ehv,Jiang:2024cqj}. However, in this paper, we primarily focus on ALPs with intermediate lifetimes, which are well-suited for efficient detection within the inner tracking detector.

Based on above discussions, we explore the possibility of detecting long-lived ALPs predicted by the photophobic scenario within the inner tracking detector at the HL-LHC with the center-of-mass energy $\sqrt{s}=14 $ TeV and integrated luminosity $\mathcal{L}=$ $3$ ab$^{-1}$. We consider the process $pp \to \gamma a$, where the ALP decays into a pair of displaced charged leptons. We find that the expected sensitivity of HL-LHC to long-lived ALPs via displaced muon signatures has already been excluded by current experiments. However, decays into displaced tau leptons remain a promising avenue. These tau leptons subsequently decay into charged particles through both leptonic and hadronic channels, leaving tracks with large transverse impact parameters in the inner tracking detector and providing distinctive experimental signatures.

This article is structured as follows. In Sec. II, we introduce the photophobic ALP framework and discuss the possible decay modes and lifetime of ALP. In Sec. III, we present a detailed analysis for the possibility of detecting long-lived ALPs  via the process $pp \to \gamma a~(a \to \tau^+ \tau^-)$ with subsequent hadronic and leptonic tau decays within the inner tracking detector at the HL-LHC. Finally, the conclusions and discussions are summarized in Sec. IV.

%%%%%%%%%%%%%%%%%%%%%%%%%%%%%%%%%%%%%%%%%%%%%%%%%%
%%%%%%%%%%%%%%%%%%%%%%%%%%%%%%%%%%%%%%%%%%%%%%%%%%
\section{The theory framework}
%%%%%%%%%%%%%%%%%%%%%%%%%%%%%%%%%%%%%%%%%%%%%%%%%%
%%%%%%%%%%%%%%%%%%%%%%%%%%%%%%%%%%%%%%%%%%%%%%%%%%

We consider the scenario in which the ALP couples exclusively to $SU(2)_L$ and $U(1)_Y$ gauge bosons, with no direct couplings to the gluons $g_{aGG}$ and fermions $g_{aff}$. This scenario is known as the photophobic ALP scenario~\cite{Craig:2018kne}, and its features are consistent with the UV boundary conditions. The the effective Lagrangian can than be written as

\begin{equation}
\begin{aligned}
\mathcal{L}_{\rm ALP} &=
\frac{1}{2}\partial_{\mu}a\partial^{\mu}a
-\frac{1}{2} m_{a}^{2} a^{2} -
C_{\tilde{W}} \frac{a}{f_a} W^{i}_{\mu\nu} \tilde{W}^{\mu\nu,i} - C_{\tilde{B}} \frac{a}{f_a} B_{\mu\nu} \tilde{B}^{\mu\nu},
\label{eq:1}
\end{aligned}
\end{equation}
where $W^{i}_{\mu\nu}$ and $B_{\mu\nu}$ are respectively field strength tensors for the gauge groups $SU(2)_{L}$ and $U(1)_{Y}$, and their dual field strength tensors are defined as $\tilde{V}^{\mu\nu}=\frac{1}{2}\epsilon^{\mu\nu\lambda\kappa}V_{\lambda\kappa}$ with $\epsilon^{\mu\nu\lambda\kappa}$ being the Levi-Civita symbol. $C_{\tilde{W}}$ and $ C_{\tilde{B}}$ denote the corresponding coupling constants. $f_a$ is the global symmetry breaking scale being independent of the ALP mass $m_a$.

After electroweak symmetry breaking, the ALP-gauge boson effective couplings presented in Eq.~(\ref{eq:1}) can be rewritten in terms of the mass eigenstates of electroweak gauge bosons,
\begin{equation}
\begin{aligned}
\mathcal{L}_{\mathrm{ALP}}
&= -\frac{1}{4}g_{a\gamma\gamma}aF_{\mu\nu}\tilde{F}^{\mu\nu}
-\frac{1}{2}g_{a\gamma Z}aZ_{\mu\nu}\tilde{F}^{\mu\nu}
-\frac{1}{4}g_{aZZ}aZ_{\mu\nu}\tilde{Z}^{\mu\nu}
- \frac{1}{2} g_{aWW} \,a\,W_{\mu \nu}^+ \tilde{W}^{\mu \nu,-},
\label{eq:2}
\end{aligned}
\end{equation}
where the coupling constants are expressed as
\begin{equation} \label{eq:3}
    \begin{aligned}
       g_{a\gamma\gamma} &= \frac{4}{f_{a}}(s_{W}^{2}c_{\tilde{W}}+c_{W}^{2}c_{\tilde{B}}), ~~~~~~~
g_{aZ\gamma} = \frac{2}{f_{a}}(c_{\tilde{W}}-c_{\tilde{B}})s_{2W}, \\
g_{aZZ} &= \frac{4}{f_{a}}(c_{W}^{2}c_{\tilde{W}}+s_{W}^{2}c_{\tilde{B}}), ~~~~~~~
g_{aWW} = \frac{4}{f_{a}}c_{\tilde{W}}.
    \end{aligned}
\end{equation}
Here $c_W \equiv \cos\theta_W$, $s_W \equiv \sin\theta_W$ and $s_{2W} \equiv \sin2\theta_W$, with $\theta_W$ being the Weinberg angle. In the photophobic ALP scenario, the coupling between ALP and diphoton is absent at tree level, satisfying the UV boundary conditions of the scenario~(see Ref.~\cite{Craig:2018kne} for details). As a result,  $g_{a\gamma\gamma}=\frac{4}{f_{a}} (s_{W}^{2} c_{\tilde{W}}+c_{W}^{2} c_{\tilde{B}}) = 0$ leads to the condition $s_{W}^{2} c_{\tilde{W}}+c_{W}^{2} c_{\tilde{B}} = 0$, which implies a proportional relationship between $c_{\tilde{W}}$ and $c_{\tilde{B}}$. Thus, the couplings of ALP with $Z\gamma$ and $ZZ$ are expressed solely in terms of $g_{aWW}$
\begin{equation}\label{eq:4}
g_{aZ\gamma} = \frac{s_{W}}{c_{W}}g_{aWW},\quad
g_{aZZ} = \frac{c_{2W}}{c_{W}^{2}}g_{aWW}.
\end{equation}

Within the photophobic ALP framework described above, there are no tree-level couplings of ALP to the fermions $g_{aff}$, photons $g_{a\gamma\gamma}$ and gluons $g_{aGG}$. However, the ALP-fermion coupling can be generated by the couplings of ALP to the electroweak gauge bosons at one loop via renormalization group evolution~(RGE), as discussed in Refs.~\cite{Bauer:2017ris,Bauer:2020jbp}. The effective ALP-fermion coupling induced at one loop is given by
\begin{equation}\label{eq:5}
\begin{aligned}
    g_{aff}^{\rm eff} &=
    \frac{3}{4s_{W}^{2}}\frac{\alpha}{4\pi}g_{aWW} \ln{\frac{\Lambda^{2}}{m_{W}^{2}}}
     \\ &\quad
    +\frac{3}{c_{W}^2}\frac{\alpha}{4\pi}g_{aWW}Q_{f}
        (I_{3}^{f}-2Q_{f}s_{W}^{2})
        \ln{\frac{\Lambda^{2}}{m_{Z}^{2}}}
     \\ &\quad
    +\frac{3 c_{2W}}{s_{W}^{2}c_{W}^{4}}\frac{\alpha}{4\pi}g_{aWW}
        (Q_{f}^{2}s_{W}^{4}-I_{3}^{f}Q_{f}s_{W}^{2}+\frac{1}{8})
        \ln{\frac{\Lambda^{2}}{m_{Z}^{2}}},
\end{aligned}
\end{equation}
where $Q_{f}$ and $I_{3}^{f}$ represent the electric charge and the third component of weak isospin of the fermion $f$, respectively. In the above expression, only the logarithmically enhanced terms $\ln(\Lambda^2 / m_V^2)$ ($V=Z, W$) are included. We take the cutoff scale $\Lambda$ to be $4 \pi f_a$ in this paper. Contributions from subleading terms have been computed in Refs.~\cite{Bauer:2017ris,Bauer:2020jbp,Bonilla:2021ufe}, and their results indicate that these contributions are extremely small. Therefore, they are not considered in our numerical calculations.

If the ALP is massive, the shift symmetry is softly broken, which allows for the generation of both the ALP-photon and ALP-gluon couplings at loop level~\cite{Bauer:2017ris}. In particular, the effective ALP-photon coupling $g^{\rm eff}_{a\gamma\gamma}$ induced by the the couplings of ALP to the electroweak gauge bosons at one loop is given by\footnote{Note that the ALP-photon coupling induced at one loop is not altered by RGE running, while the ALP-photon coupling induced by the effective ALP-fermion coupling is altered by RGE running. However, the impact is minimal.}
\begin{equation}\label{eq:6}
    g_{a\gamma\gamma}^{\rm eff} = \frac{2\alpha}{\pi} g_{aWW}B_{2}(\tau_{W}),
\end{equation}
where $\alpha = e^2/(4\pi)$ is the fine-structure constant with $e = g s_{W}$ being the QED coupling.
The loop function $B_{2}$ is defined as
\begin{equation}\label{eq:7}
    B_{2}(\tau) = 1-(\tau-1)f^{2}(\tau),\quad \text{with} f(\tau) =
    \begin{cases}
        \arcsin{\frac{1}{\sqrt{\tau}}}, & \text{for } \tau \geq 1 \\
        \frac{\pi}{2} + \frac{i}{2} \log{\frac{1+\sqrt{1-\tau}}{1-\sqrt{1-\tau}}}, & \text{for } \tau < 1
    \end{cases},
\end{equation}
where $\tau_{W}=4m_{W}^{2}/m_{a}^{2}$. It should be noted that the ALP-gluon coupling can be generated via the effective ALP-fermion couplings. However, these effective ALP-fermion couplings are induced by the the couplings of ALP to the electroweak gauge bosons at one loop. Therefore, the ALP-gluon coupling indeed arises only at two-loop level and is strongly suppressed. While this coupling is slightly altered by RGE running, its contribution to the total ALP decay width is very small and can be safely neglected in this study.

From above discussions, we can see that the ALP can decay to electroweak gauge bosons~($WW$, $ZZ$, and $Z\gamma$) at tree level, whereas decays into photons and fermions only occur at the loop level. The corresponding decay widths are given by

\begin{equation}\label{eq:8}
\begin{aligned}
 &
 \Gamma(a\to\gamma\gamma) =  \frac{m_a^3}{64\pi}|g_{a\gamma\gamma}^\text{eff}|^2,
 \\
 &
  \Gamma(a\to f \bar{f}) =  N_{c}^f \frac{m_a m_{f}^2|g_{aff}^\text{eff}|^2}{8\pi}\sqrt{1-\frac{4m_{f}^2}{m_a^2}},
 \\
 &
  \Gamma(a\to Z \gamma) =  \frac{m_a^3 s_{W}^2}{32\pi c_{W}^2}|g_{aWW}|^2 (1-\frac{m_Z^2}{m_a^2})^3,
  \\
&
  \Gamma(a\to Z Z) =  \frac{m_a^3 c_{2W}^2}{64\pi c_{W}^4}|g_{aWW}|^2 (1-4\frac{m_Z^2}{m_a^2})^{3/2},
  \\
&
  \Gamma(a\to W^{+} W^{-}) =  \frac{m_a^3}{32\pi}|g_{a W W}|^2 (1-4\frac{m_W^2}{m_a^2})^{3/2}.
\end{aligned}
\end{equation}
Here $N_c^f$ represents the color factor of the fermion $f$, its values taken 1 for leptons and 3 for quarks. As shown in Eq.~\eqref{eq:8}, the decay widths are sensitive to the ALP mass $m_a$ and its couplings. All the couplings in the above expression can be expressed in terms of the coupling $g_{aWW}$, as shown in Eqs.~\eqref{eq:4}-\eqref{eq:6}. Consequently, the ALP phenomenology considered in this paper can be simplified to depend solely on the two parameters, the ALP mass $m_a$ and the coupling $g_{aWW}$. In our numerical calculations, we choose $\Lambda = 4\pi f_a = 4\pi \times 10$ TeV to evaluate the loop-induced ALP decays.

\begin{figure}[H]
\begin{center}
\includegraphics [scale=0.35] {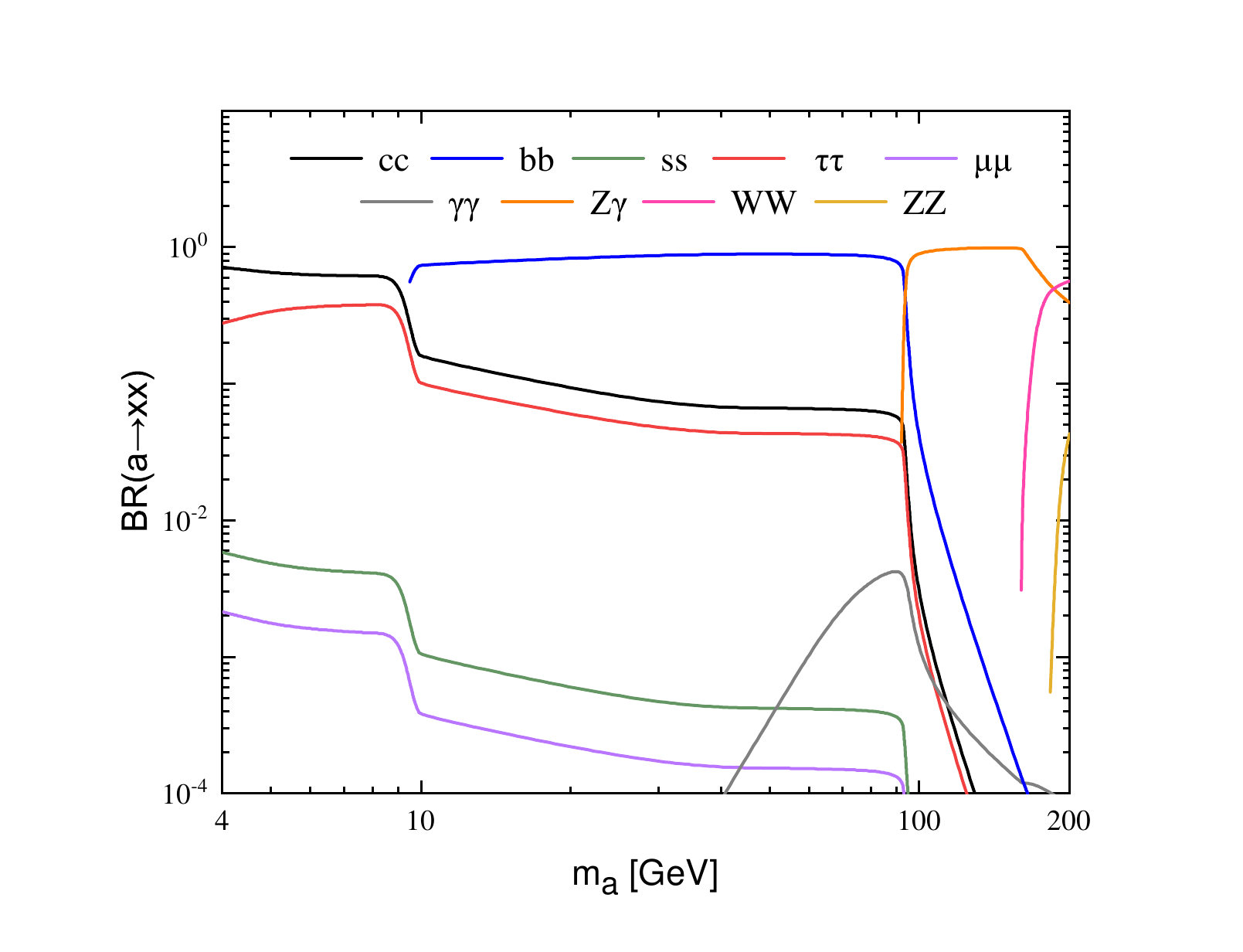}
\caption{The branching ratios for different photophobic ALP decay modes as functions of its mass $m_a$.}
\label{fig:1}
\end{center}
\end{figure}

\begin{figure}[H]
\begin{center}
\includegraphics [scale=0.35] {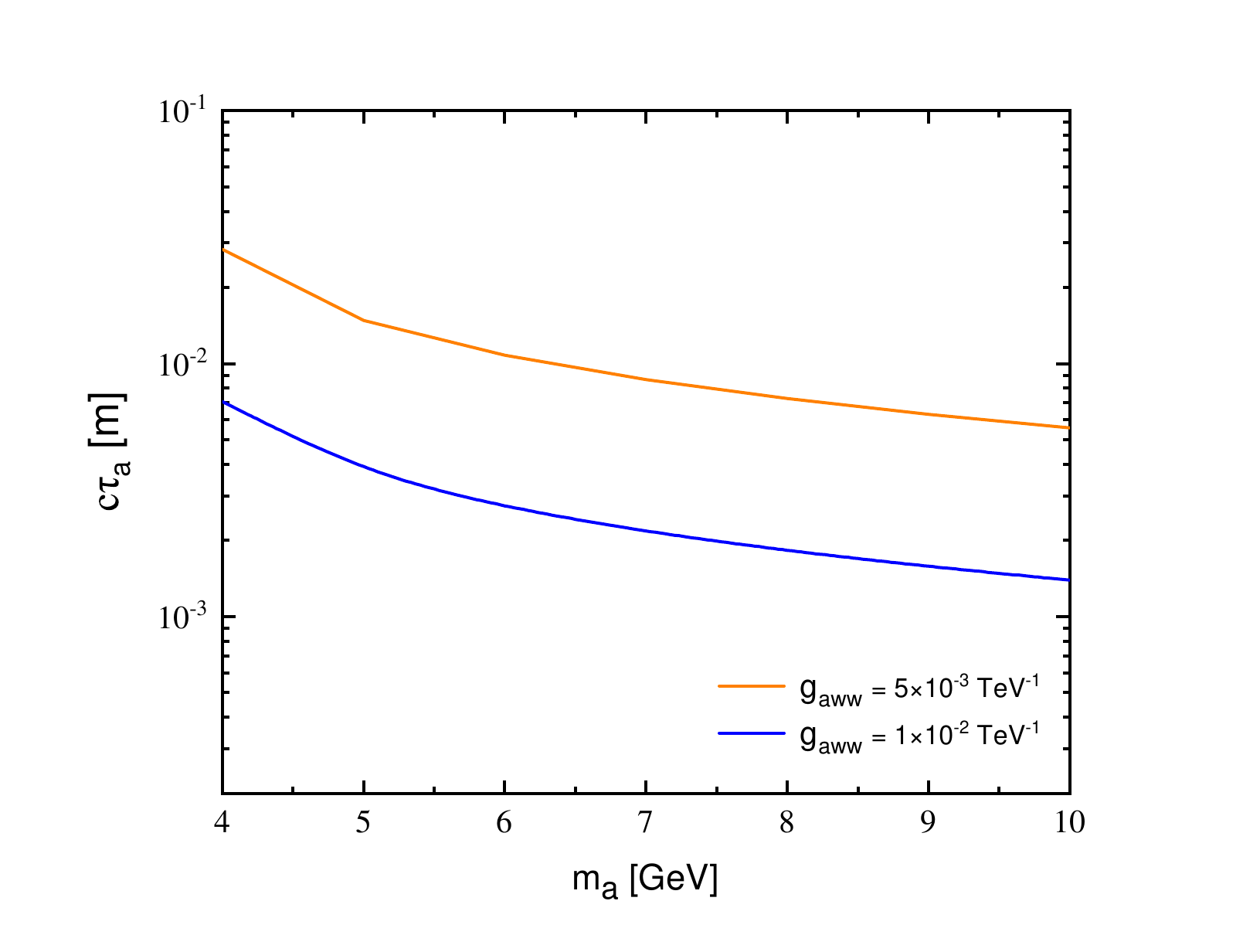}
\caption{The proper decay length of the ALP as a function of its mass $m_a$ for different values of the  coupling $g_{aWW}$.}
\label{fig:2}
\end{center}
\end{figure}

The specific branching fractions for different ALP decay modes mentioned above are shown in Fig.~\ref{fig:1}. As illustrated in this figure, for ALPs with relatively low masses, tree-level decays into massive electroweak gauge bosons~($W$ and $Z$) are kinematically forbidden. In this mass regime, the ALP predominantly decays into fermions and photons at one loop. Among these kinematically allowed channels, decays into light charged leptons are suppressed due to their small masses, while the diphoton channel is further suppressed owing to the much weaker loop-induced ALP-photon coupling compared to the ALP-lepton couplings. Given that the LHCb collaboration has already provided stringent limits on light long-lived ALPs with masses below about $4$ GeV through rare B meson decays~\cite{LHCb:2016awg,LHCb:2015nkv}, we consider heavier ALPs with masses in the range of $4$ to $10$ GeV for DV signature analysis in the context of the photophobic ALP model.\footnote{It should be noted that in the mass range considered in this paper, there exist pseudoscalar SM resonances that could potentially mix with the ALP $a$ (see, e.g., Ref.~\cite{Haisch:2018kqx}). However, in the photophobic ALP scenario, the couplings of ALP to the SM fermions arise only at one loop and are therefore strongly suppressed. As a result, the impact of such mixing on the ALP decays is expected to be small and is therefore neglected in this study.} Based on the above discussion, the dominant decay modes in this mass range are $\tau^+ \tau^-$ and $c \bar{c}$, as shown in Fig.~\ref{fig:1}. Then its total decay width can be written as $\Gamma_a= \Gamma(a\to\tau^+ \tau^- )+ \Gamma(a\to c \bar{c})$. The observed decay length of the ALP in the laboratory frame depends on both its proper decay length~(or proper lifetime) $c \tau_a$, velocity $\beta_a$ as well as Lorentz boost factors $\gamma_a$, which is given by
\begin{equation}\label{eq:9}
   L_a = \beta_a \gamma_a c \tau_a = \frac{c|p_a|}{m_a \Gamma_a}.
\end{equation}
Where $\tau_a = 1/\Gamma_a$, $p_a$ is the momentum of ALP and c is the speed of light. The proper decay length $c \tau_a$ as function of the ALP mass $m_a$ for different values of the coupling $g_{aWW}$ is shown in Fig.~\ref{fig:2}. It is evident that the proper decay length varies significantly with different values of $m_a$ and $g_{aWW}$, increasing with the decrease of both
$m_a$ and $g_{aWW}$. This wide range of proper decay lengths suggests that long-lived ALPs can be observed over a broad parameter space, which can be effectively probed within the inner tracking detector at the HL-LHC. In addition, long-lived ALPs can also be explored in other sub-detectors, such as the ECAL, HCAL and MS, as discussed in Refs.~\cite{ATLAS:2023ian,Carmona:2022jid,ATLAS:2024ocv,Mitridate:2023tbj}. However, the analysis of displaced calorimeter deposits in the ECAL and HCAL is not considered in this study, as it requires dedicated simulations and appropriate trigger designs. Additionally, the analysis in the MS is excluded due to its greater distance from the interaction point, which makes it less sensitive to short-lived displaced decays compared to the ID.

We consider the process $pp \to \gamma a$ for DV searches, where the ALP predominantly decays into a pair of tau leptons or charm quarks. Although the charm channel has a relatively larger branching ratio, it is hindered by significant QCD backgrounds and challenges in reconstructing displaced charm decays with sufficient precision.
Hence, we focus on the $\tau^+ \tau^-$ final state as the primary signature for DV searches. Since tau leptons are unstable, they decay either leptonically into electrons or muons with missing transverse energy or hadronically into mesons with missing transverse energy. For a comprehensive analysis, we simulate the subsequent hadronic and leptonic decays of these tau leptons.

%%%%%%%%%%%%%%%%%%%%%%%%%%%%%%%%%%%%%%%%%%%%%%%%%%
%%%%%%%%%%%%%%%%%%%%%%%%%%%%%%%%%%%%%%%%%%%%%%%%%%
\section{The possibility of detecting long-lived axion-like particles at the HL-LHC}
%%%%%%%%%%%%%%%%%%%%%%%%%%%%%%%%%%%%%%%%%%%%%%%%%%
%%%%%%%%%%%%%%%%%%%%%%%%%%%%%%%%%%%%%%%%%%%%%%%%%%

\begin{figure}[H]
\begin{center}
\subfigure[]{\includegraphics [scale=0.55] {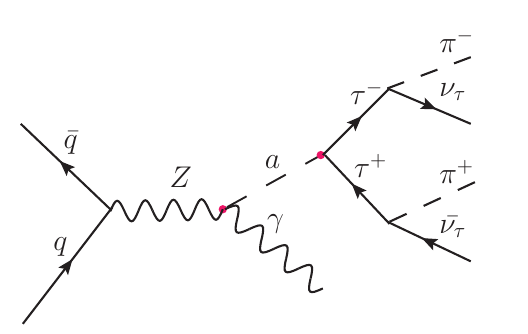}}
\hspace{0.3in}
\subfigure[]{\includegraphics [scale=0.55] {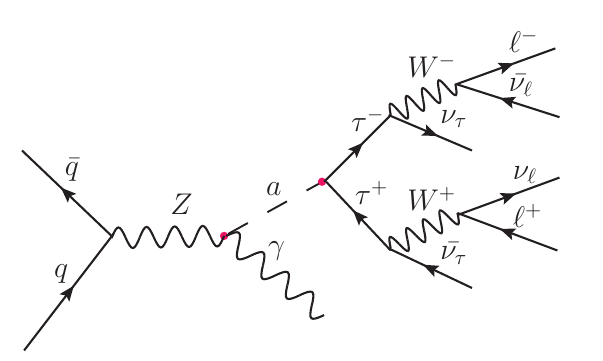}}
\caption{The Feynman diagrams for the process $pp \to \gamma a~(a \to \tau^+ \tau^-)$ with subsequent hadronic tau decay $ \tau^+ \tau^- \to \pi^+ \pi^- \slashed{E}_T $~(a) and leptonic tau decay $ \tau^+ \tau^- \to \ell^+ \ell^- \slashed{E}_T $~(b).}
\label{fig:3}
\end{center}
\end{figure}

We now investigate the potential for detecting long-lived ALPs with masses in the range $4 \, \text{GeV} \leq m_a \leq 10 \, \text{GeV}$ predicted by the photophobic scenario within the inner tracking detector at the HL-LHC with $\sqrt{s} = 14$ TeV and $\mathcal{L} = 3$ ab$^{-1}$. We consider the process $pp \to \gamma a$ with $a \to \tau^+ \tau^-$, where the tau leptons further decay either hadronically as $\tau^+ \tau^- \to \pi^+ \pi^- \slashed{E}_T$ or leptonically as $\tau^+ \tau^- \to \ell^+ \ell^- \slashed{E}_T$~($\ell = e, \mu$), leading to two distinct signals: $\pi^+ \pi^- \gamma \slashed{E}_T $ and $\ell^+ \ell^- \gamma \slashed{E}_T $. The corresponding Feynman diagrams are shown in Fig.~\ref{fig:3}.

We employ \verb"FeynRules" package~\cite{Alloul:2013bka} to generate the model file for the effective Lagrangian. A Monte Carlo simulation is carried out to investigate the potential for detecting long-lived ALPs within the inner tracking detector at the HL-LHC. The signal and background events discussed in the following subsections are generated by \verb"MadGraph5_aMC@NLO"~\cite{Alwall:2014hca} with the \verb"TauDecay" package~\cite{Hagiwara:2012vz} and then passed to \verb"PYTHIA8"~\cite{Sjostrand:2014zea} for parton showering and hadronization. At the generator level, we applied the following basic cuts for the signal and SM background events:
\begin{equation}\label{eq:10}
\begin{split}
\hspace{-3em} \text{For leptonic tau decay process}: ~~& p_T^l > 10~\text{GeV}, \quad |\eta_l| < 2.5,~\text{with}~\ell = e, \mu, \\
    & p_T^\gamma > 10~\text{GeV}, \quad |\eta_\gamma| < 2.5, \\
    & \Delta R_{ll} > 0.4, \quad \Delta R_{l\gamma} > 0.4.\\
\hspace{-3em} \text{For hadronic tau decay process}: ~~ & p_T^\gamma > 10~\text{GeV}, \quad |\eta_\gamma| < 2.5.
\end{split}
\end{equation}
Where $p_T^l$ and $p_T^\gamma$ denote the transverse momentum of leptons and photons, $|\eta_l|$ and $|\eta_\gamma|$ represent the absolute values of the pseudorapidity of leptons and photons, $\Delta R_{ll}$ and $\Delta R_{l\gamma}$ refer to the angular separation between leptons, and between photon and lepton, in which $\Delta R$ is defined as $\sqrt{(\Delta \phi)^{2}+(\Delta \eta)^{2}}$. For all event generations, we use the \texttt{NN23LO1} parton distribution functions (PDFs)~\cite{Ball:2012cx,Deans:2013mha}. The fast detector simulation is performed using the simplified fast simulation (SFS) framework~\cite{Araz:2020lnp} embedded in \verb"MadAnalysis5"~\cite{Conte:2012fm,Conte:2014zja,Conte:2018vmg}. The detector configuration is based on the SFS card provided in the CMS-EXO-16-022 analysis template~\cite{Araz:2020lnp}. Kinematic and cut-based analysis are performed using \texttt{C++} programming in the expert mode of \verb"MadAnalysis5".

\subsection{Searching for long-lived ALPs via the $ \pi^+ \pi^- \gamma \slashed{E}_T $ signal }\label{subsec1}

Now let's consider the signal from $pp \to \gamma a \to \gamma \tau^+ \tau^-$ with subsequent hadronic tau decays at the $14$ TeV HL-LHC with $\mathcal{L}=$ $3$ ab$^{-1}$, which is characterized by a pair of charged pions, a photon and missing transverse energy. Its production cross sections at the HL-LHC decrease slowly with $m_a$ increasing, while showing a clear enhancement as the coupling $g_{aWW}$ increases. For the ALPs with masses in the range of $4-10$ GeV, the values of the production cross sections vary from $8.88 \times 10^{-6}$ to $8.66 \times 10^{-6}$ pb and from $2.22 \times 10^{-6}$ to $2.16 \times 10^{-6}$ pb for $g_{aWW}$ equaling to $1 \times 10^{-2}$ TeV$^{-1}$ and $5 \times 10^{-3}$ TeV$^{-1}$, respectively. These results have been obtained after applying the basic cuts described earlier.
The SM background mainly arises from the Drell-Yan process $pp \to \gamma^*/Z^* \to \tau^+ \tau^-$, where one of the $\tau$ leptons radiates a photon before decaying hadronically. The subdominant contribution to the SM background originates from the process $pp \to \gamma \tau^+ \tau^-$, where the photon is radiated from an initial-state jet and both $\tau$ leptons decay hadronically  $\tau^{\pm} \to \pi^{\pm} \nu_{\tau}$. For the SM background, the value of the total production cross section is about $0.28$ pb. In this analysis, potential instrumental backgrounds mainly arise from random-track crossings, vertices coming from dense detector regions and pile-up. The events from random-track crossings can be estimated by using data-driven methods such as event mixing and track flipping. For example, reference~\cite{ATLAS:2019fwx} employs these methods to estimate the number of events, which is about $\mathcal{O}(10^{-3})$ and negligible. Vertices from dense detector regions can be effectively vetoed by using detailed material maps~\cite{Alimena:2019zri,ATLAS:2017tny}. Additionally, pile-up background can be suppressed by reconstructing a merged vertex from the decays of tau leptons, as detailed in Ref.~\cite{Shan:2024pcc}. A complete quantitative assessment of these instrumental backgrounds requires the application of specialized technical frameworks, which are beyond the scope of this study. Therefore, the results presented in this paper are based on the idealized assumption that the contributions of these instrumental backgrounds are negligible. A systematic evaluation of these backgrounds will be considered in future work.

\begin{figure}[H]
\begin{center}
\subfigure[]{\includegraphics [scale=0.35] {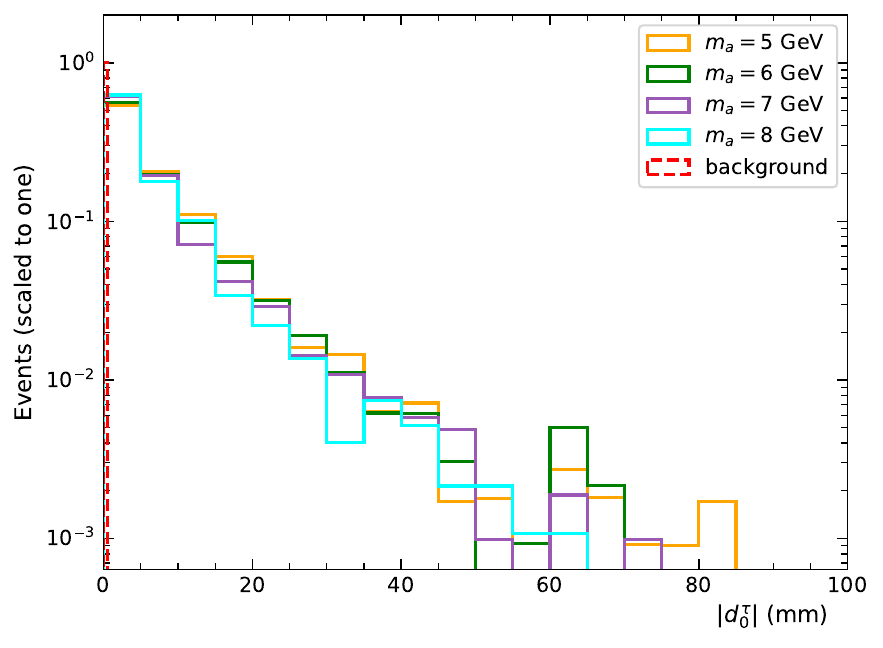}}
\hspace{0.1in}
\subfigure[]{\includegraphics [scale=0.35] {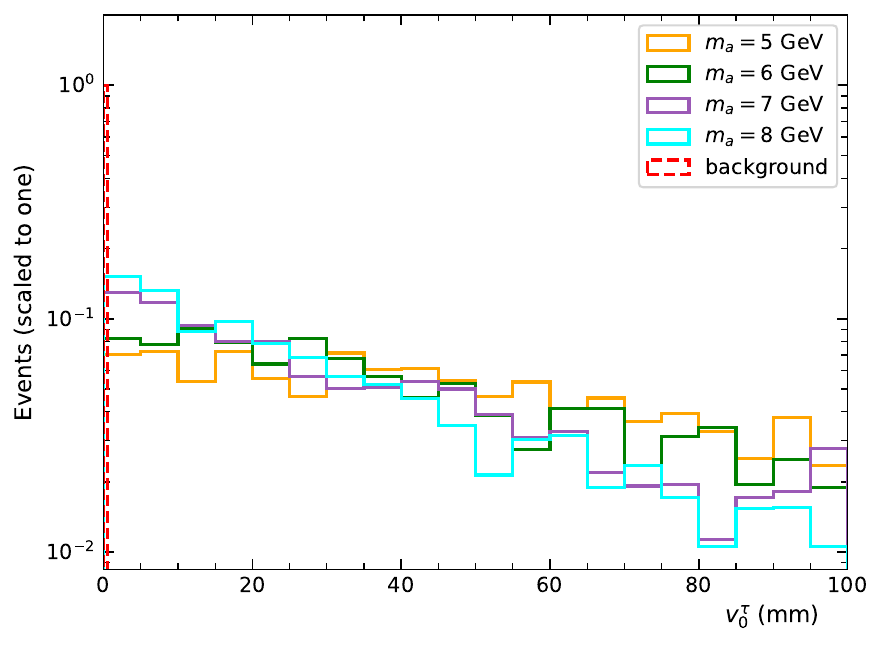}}
\hspace{0.1in}
\subfigure[]{\includegraphics [scale=0.35] {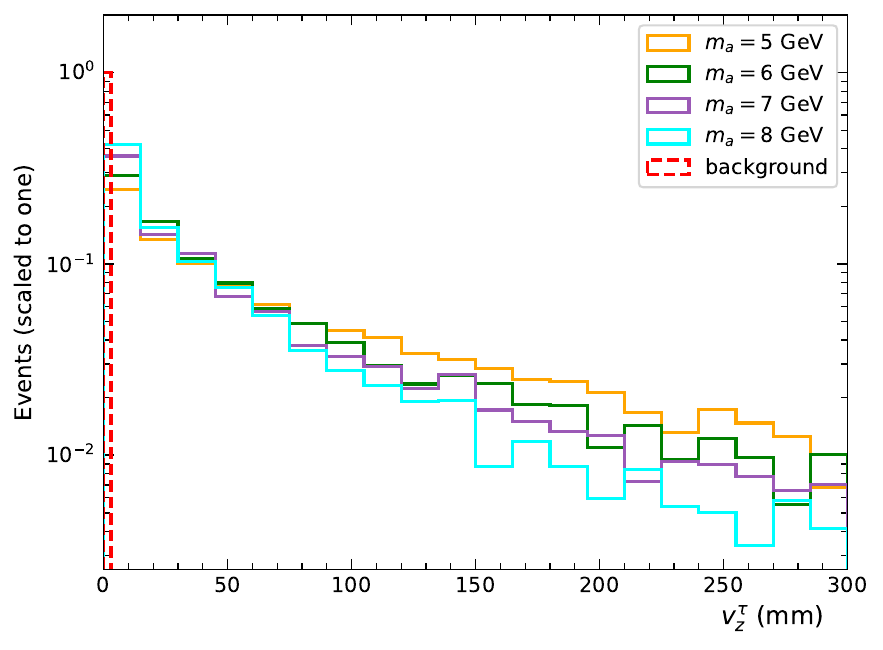}}
\subfigure[]{\includegraphics [scale=0.35] {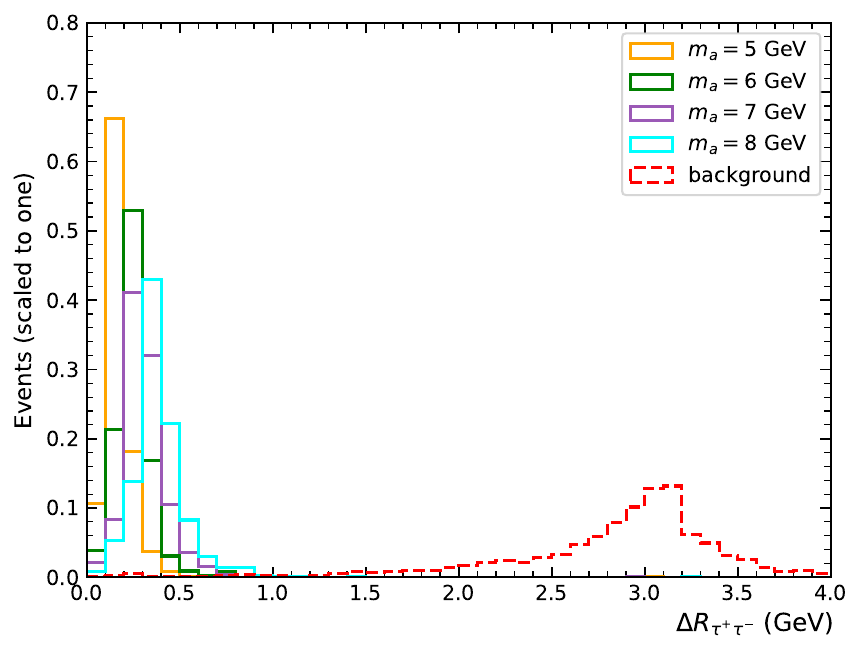}}
\subfigure[]{\includegraphics [scale=0.35] {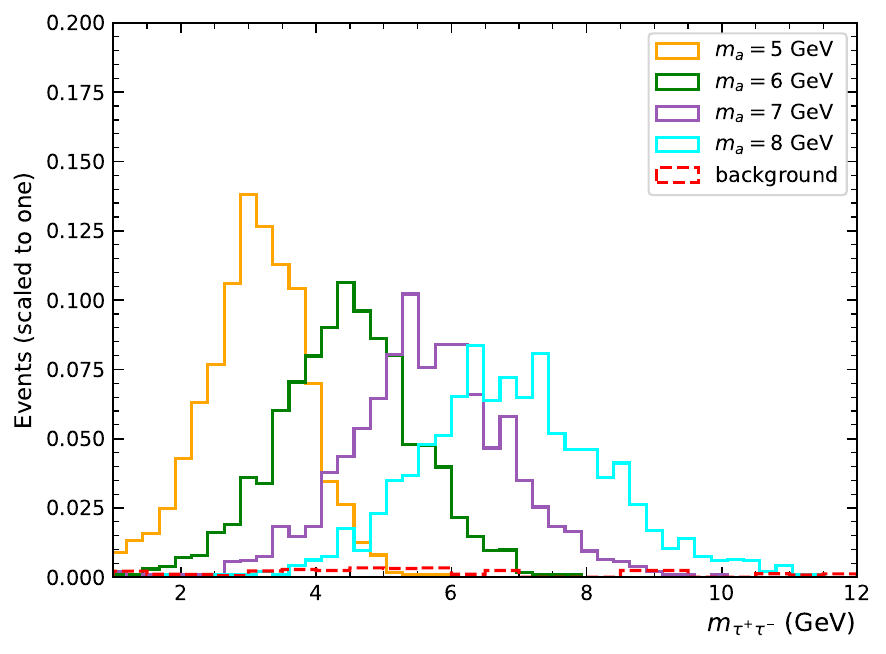}}
\caption{The normalized distributions of the observables $|d_0^{\tau}|$, $v_0^{\tau}$, $v_z^{\tau}$, $\Delta R_{\tau^+\tau^-}$ and $m_{\tau^+\tau^-}$ for the signal of selected ALP-mass benchmark points and the SM background for $g_{aWW}=5 \times 10^{-3}$ TeV$^{-1}$ at the 14 TeV HL-LHC with $\mathcal{L}=$ $3$ ab$^{-1}$.}
\label{fig:4}
\end{center}
\end{figure}

In our analysis, hadronic tau decays are reconstructed as narrow $\tau$ jets, each associated with a single charged track. To retain a sufficient number of signal events, a pre-selection is first applied, requiring $\tau$ jet with the transverse momentum $p_T^{\tau} > 10$ GeV and the pseudorapidity $|\eta_{\tau}| < 2.4$, along with at least one pair of oppositely charged $\tau$ jets. To effectively distinguish the signal from the SM background, we perform a cut-based analysis on several crucial kinematic observables. We analyze the transverse impact parameter $d_0^{\tau}$ of the charged track associated with the $\tau$ jet, as well as the transverse distance $v_0^{\tau}$ and longitudinal distance $v_z^{\tau}$, which represent the distance from the reconstructed $\tau$ jet production vertex to the interaction point. To further explore the kinematics of the reconstructed hadronic $\tau$ jets, we also analyze the angular separation $\Delta R_{\tau^+\tau^-}$ and the invariant mass $m_{\tau^+\tau^-}$ of the $\tau$-jet pair. The normalized distributions of these kinematic variables at the 14 TeV HL-LHC with $\mathcal{L}=$ $3$ ab$^{-1}$ are shown in Fig.~\ref{fig:4}, where the solid lines stand for signal events and dashed lines stand for background events. The benchmark ALP mass points $m_a = 5$, $6$, $7$ and $8$ GeV are selected with $g_{aWW}=5 \times 10^{-3}$ TeV$^{-1}$.

As shown in Fig.~\ref{fig:4}, for the signal events, the long-lived ALP travels a measurable distance before decaying into a pair of $\tau$ leptons, leading to larger values of $|d_0^{\tau}|$, $v_0^{\tau}$ and $v_z^{\tau}$. In contrast, the $\tau^+ \tau^-$ for the SM background are produced promptly at the interaction point, yielding minimal values for these displacement variables. Additionally, the angular separation $\Delta R_{\tau^+\tau^-}$ exhibits a clear distinction between the signal and background events. For the signal events, the two reconstructed $\tau$ jets original from hadronic tau decays produced by the light ALP. As a result, the two $\tau$ jets are more collimated and have a smaller angular separation. In contrast, the background events typically show a broader distribution of angular separation between the two $\tau$ jets. It is also evident that the signal and background can be well distinguished in the invariant mass $m_{\tau^+\tau^-}$ distribution. The signal events exhibit a distinct peak in the invariant mass distribution, located slightly below the corresponding ALP mass due to the energy carried away by neutrinos from the tau decays. As the ALP mass increases, the peak systematically moves to higher values and broadens slightly. While the SM background events do not show a distinct peak and have a more spread-out invariant mass distribution.

Based on the characteristic features of the kinematic distributions discussed above, optimized cuts presented in Table~\ref{tab1} are further applied to suppress the SM background and enhance the statistical significance of the signal. To ensure that the ALP decays occur within the inner tracking detector, we require that these parameters $|d_0^{\tau}|$, $v_0^{\tau}$ and $v_z^{\tau}$ respectively satisfy the conditions $0.2 \, \text{mm} < |d_0^{\tau}| < 100 \, \text{mm}$, $v_0^{\tau} < 10 \, \text{cm}$ and $v_z^{\tau} < 30 \, \text{cm}$, which are adapted from Refs.~\cite{Araz:2021akd,Lu:2024ade} and also summarized in Table~\ref{tab1}. With these optimized cuts, particularly the requirements on $|d_0^{\tau}|$, $v_0^{\tau}$ and $v_z^{\tau}$, the SM background is expected to be highly suppressed and therefore considered negligible in our analysis. The cumulative selection efficiency $\epsilon$ and production cross sections of the $\pi^+ \pi^- \gamma \slashed{E}_T $ signal after imposing improved cuts for several ALP mass benchmark points and the coupling $g_{aWW}=5\times10^{-3}$ TeV$^{-1}$ at the $14$ TeV HL-LHC with $\mathcal{L}=$ $3$ ab$^{-1}$ are given in Table~\ref{tab2}. The values of $\epsilon$ can reach approximately $9.69\times10^{-2}$ $\sim$ $1.48\times10^{-1}$ for $m_a$ between 5 and 8 GeV.

\begin{table}[H]
\begin{center}
\setlength{\tabcolsep}{3mm}{
\caption{The optimized cuts on the $\pi^+ \pi^- \gamma \slashed{E}_T$ signal and SM background for $4~\mathrm{GeV} \leq m_a \leq 10~\mathrm{GeV}$ at the 14 TeV HL-LHC with $\mathcal{L}=$ $3$ ab$^{-1}$.}
\label{tab1}
\resizebox{0.68\textwidth}{!}{
\begin{tabular}
[c]{c|c c c c c}\hline \hline
\multicolumn{1}{c|}{\textbf{Optimized cuts}} & \textbf{Selection criteria} \\
\hline
\makecell[l]{Pre-secetion: the transverse momentum, pseudorapidity and\\
                multiplicity of the $\tau$ jet in the final states } &
\makecell[c]{$p_T^{\tau} > 10$ GeV, $|\eta_{\tau}| < 2.4$ \\$N_{\tau^{+}} \geq 1$, $N_{\tau^{-}} \geq 1$} \\
\makecell[l]{Cut 1: the transverse impact parameter} &
\makecell[c]{$0.2~\mathrm{mm} < |d_{\textnormal{0}}^{\tau}| < 100~\mathrm{mm}$} \\
\makecell[l]{Cut 2: the transverse and longitudinal distance} &
\makecell[c]{$v_{\textnormal{0}}^{\tau} < 10~\mathrm{cm}$, $v_z^{\tau} < 30~\mathrm{cm}$} \\
\makecell[l]{Cut 3:the angular separation between the two $\tau$ jets} &
\makecell[c]{$\Delta R_{\tau^+\tau^-} < 1.1$} \\
\makecell[l]{Cut 4:the invariant-mass window cut on the $\tau$-jet pair} &
\makecell[c]{$m_{\tau^{+}\tau^{-}} \leq 11~\mathrm{GeV}$} \\
\hline \hline
\end{tabular}}}
\end{center}
\end{table}

\begin{table}[H]\tiny
	\centering{
\caption{The cumulative selection efficiency $\epsilon$ and production cross section $\sigma$ of the $ \pi^+ \pi^- \gamma \slashed{E}_T $ signal after the optimized cuts applied at the $14$ TeV HL-LHC with $\mathcal{L}=$ $3$ ab$^{-1}$ for $g_{aWW}=5\times10^{-3}$ TeV$^{-1}$ with the benchmark points $m_a = 5$, $6$, $7$ and $8$ GeV.$~$\label{tab2}}
		\newcolumntype{C}[1]{>{\centering\let\newline\\\arraybackslash\hspace{50pt}}m{#1}}
		\begin{tabular}{m{2cm}<{\centering}|m{2.3cm}<{\centering} m{2.3cm}<{\centering} m{2.3cm}<{\centering}  m{2.3cm}<{\centering} m{2.3cm}<{\centering}}\hline \hline
      \multirow{2}{*}{Cuts} & \multicolumn{4}{c}{Cross sections~[pb] (selection efficiencies) for the signal}\\
     \cline{2-5}
     & $m_a=5$ GeV  & $m_a=6$ GeV  & $m_a=7$ GeV  & $m_a=8$ GeV  \\ \hline
     Generator:  & \makecell{$2.22\times10^{-6}$} & \makecell{$2.21\times10^{-6}$} & \makecell{$2.20\times10^{-6}$} & \makecell{$2.20\times10^{-6}$} \\ \hline
     Pre-secetion:  & \makecell{$4.77\times10^{-7}$\\($2.16\times10^{-1}$)} & \makecell{$4.54\times10^{-7}$\\($2.06\times10^{-1}$)} & \makecell{$4.35\times10^{-7}$\\($1.98\times10^{-1}$)} & \makecell{$4.16\times10^{-7}$\\($1.90\times10^{-1}$)} \\
     Cut 1:  & \makecell{$4.24\times10^{-7}$\\($1.92\times10^{-1}$)} & \makecell{$4.04\times10^{-7}$\\($1.84\times10^{-1}$)} & \makecell{$3.80\times10^{-7}$\\($1.73\times10^{-1}$)} & \makecell{$3.66\times10^{-7}$\\($1.67\times10^{-1}$)} \\
     Cut 2:  & \makecell{$2.14\times10^{-7}$\\($9.69\times10^{-2}$)} & \makecell{$2.70\times10^{-7}$\\($1.23\times10^{-1}$)} & \makecell{$3.06\times10^{-7}$\\($1.40\times10^{-1}$)} & \makecell{$3.27\times10^{-7}$\\($1.50\times10^{-1}$)} \\
     Cut 3:  & \makecell{$2.14\times10^{-7}$\\($9.69\times10^{-2}$)} & \makecell{$2.70\times10^{-7}$\\($1.23\times10^{-1}$)} & \makecell{$3.06\times10^{-7}$\\($1.40\times10^{-1}$)} & \makecell{$3.27\times10^{-7}$\\($1.49\times10^{-1}$)} \\
     Cut 4:  & \makecell{$2.14\times10^{-7}$\\($9.69\times10^{-2}$)} & \makecell{$2.70\times10^{-7}$\\($1.23\times10^{-1}$)} & \makecell{$3.06\times10^{-7}$\\($1.40\times10^{-1}$)} & \makecell{$3.24\times10^{-7}$\\($1.48\times10^{-1}$)} \\
     \hline \hline
	\end{tabular}}	
\end{table}

Then, the number of signal events meeting the selection criteria mentioned above is given by
\begin{equation}\label{eq:11}
\begin{split}
    N_{sig} = \epsilon \times \sigma \times \mathcal{L}.
\end{split}
\end{equation}
In our analysis, the expected number of background events is zero. Therefore, when presenting numerical results, we set the threshold for the required number of signal events to $N_{sig} \geq 3$, which precisely corresponds to the exclusion limit at $95\%$ confidence level~(C.L.). We then apply the same search strategy over wider ranges of $m_a$ and $g_{aWW}$ and derive the expected $95\%$ C.L. sensitivity of the 14 TeV HL-LHC with $\mathcal{L}=$ $3$ ab$^{-1}$ to long-lived ALPs, as shown in the orange region of the Fig.~\ref{fig:6}.

\subsection{Searching for long-lived ALPs via the $\ell^+ \ell^- \gamma \slashed{E}_T $ signal}\label{subsec2}

In this subsection, we explore the signal from the process $pp \to \gamma a \to \gamma \tau^+ \tau^-$ with subsequent leptonic tau decays, which is composed of two oppositely charged leptons $e$ or $\mu$, a photon and missing transverse energy. For ALPs with masses between $4$ and $10$ GeV, the production cross sections are in the ranges from $9.65 \times 10^{-5}$ to $9.38 \times 10^{-5}$ pb and from $2.41 \times 10^{-5}$ to $2.35 \times 10^{-5}$ pb for $g_{aWW}$ equaling to $1 \times 10^{-2}$ TeV$^{-1}$ and $5 \times 10^{-3}$ TeV$^{-1}$, respectively. Notably, the production cross sections of the $\ell^+ \ell^- \gamma \slashed{E}_T$ signal are approximately one order of magnitude larger than those of the $ \pi^+ \pi^- \gamma \slashed{E}_T $ signal. The dominant SM background arises from the Drell-Yan process $pp \to \gamma^*/Z^* \to \tau^+ \tau^-$, where one of $\tau$ leptons radiates a photon before decaying leptonically. Subleading contributions come from the process $pp \to \gamma \tau^+ \tau^-$, where the photon is emitted from an initial-state jet and both $\tau$ leptons decay leptonically. The total production cross section of the SM background is about $3.01$ pb at the $14$ TeV HL-LHC. \footnote{Additionally, long-lived decays of the SM particles, such as semi-leptonic $B$-decays, may contribute to the SM background. Displaced tracks from $B$-meson decays typically exhibit transverse impact parameters $d_0$ smaller than 2 mm~\cite{Alimena:2019zri}. Applying a $|d_0| > 2$ mm cut can effectively suppress this background. For other long-lived decays of the SM particles, we can similarly apply an optimized $d_0$ cut to mitigate the background. However, such cuts may also affect the signal, which will be further considered in future study.}

\begin{figure}[H]
\begin{center}
\subfigure[]{\includegraphics [scale=0.35] {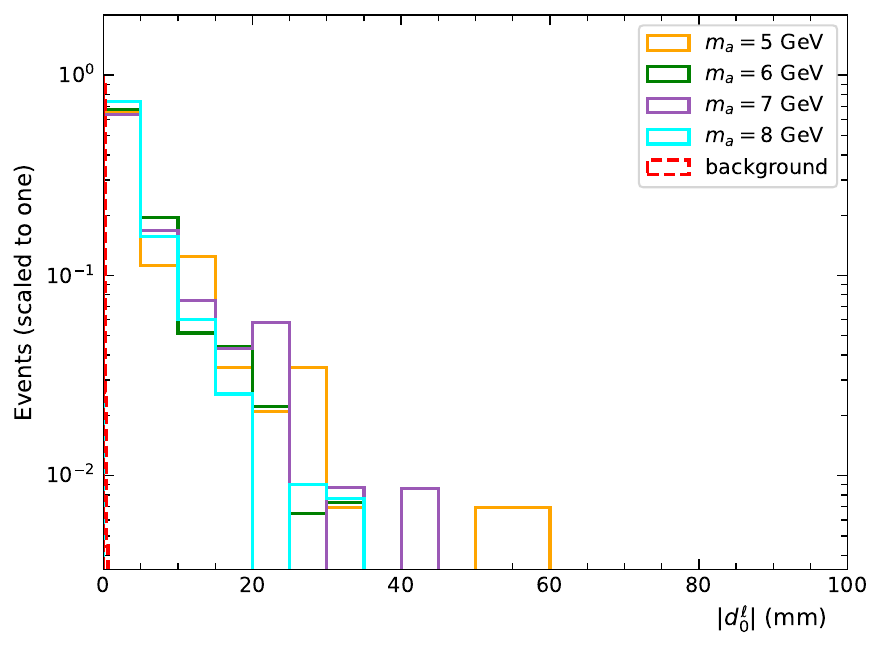}}
\hspace{0.1in}
\subfigure[]{\includegraphics [scale=0.35] {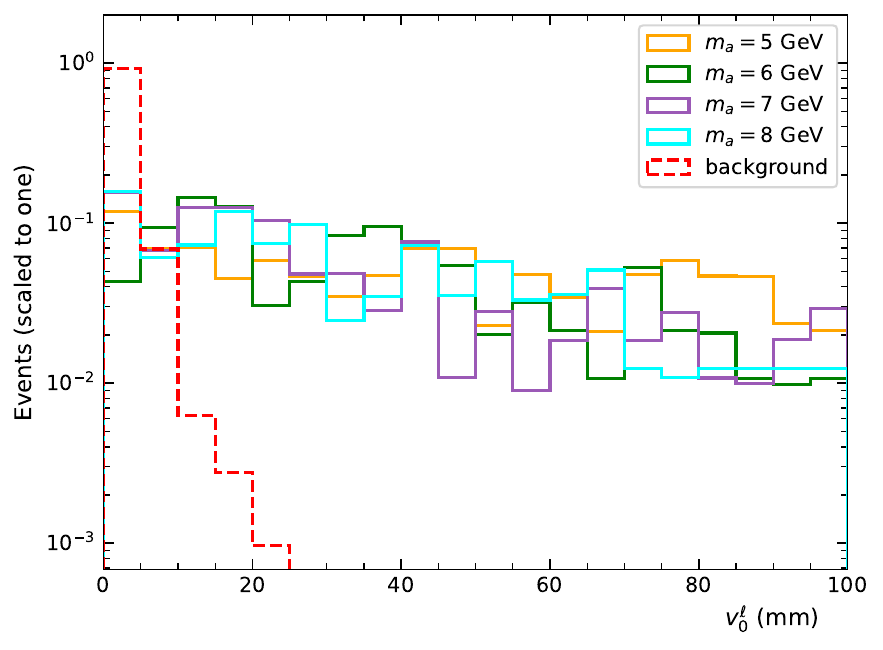}}
\hspace{0.1in}
\subfigure[]{\includegraphics [scale=0.35] {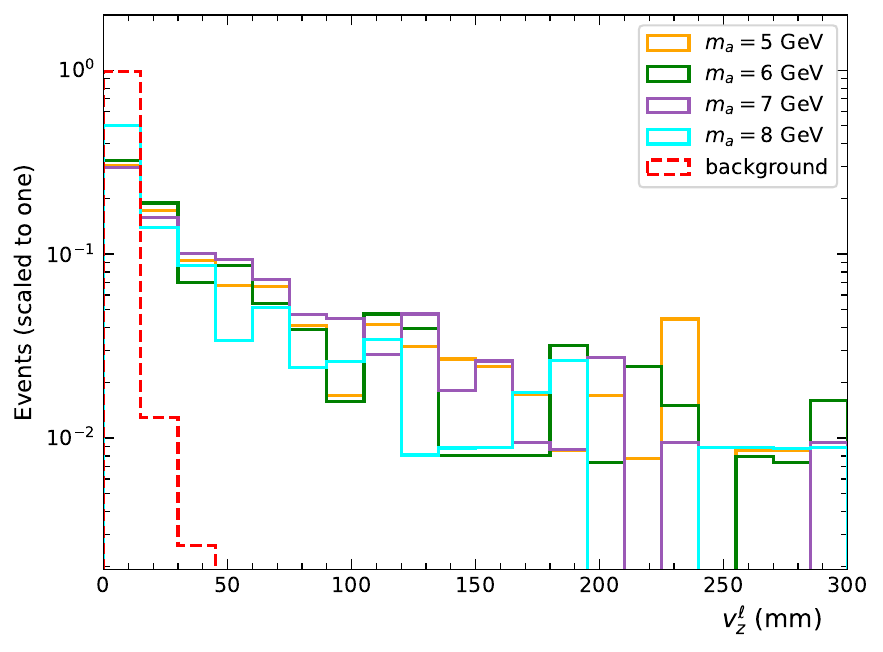}}
\subfigure[]{\includegraphics [scale=0.35] {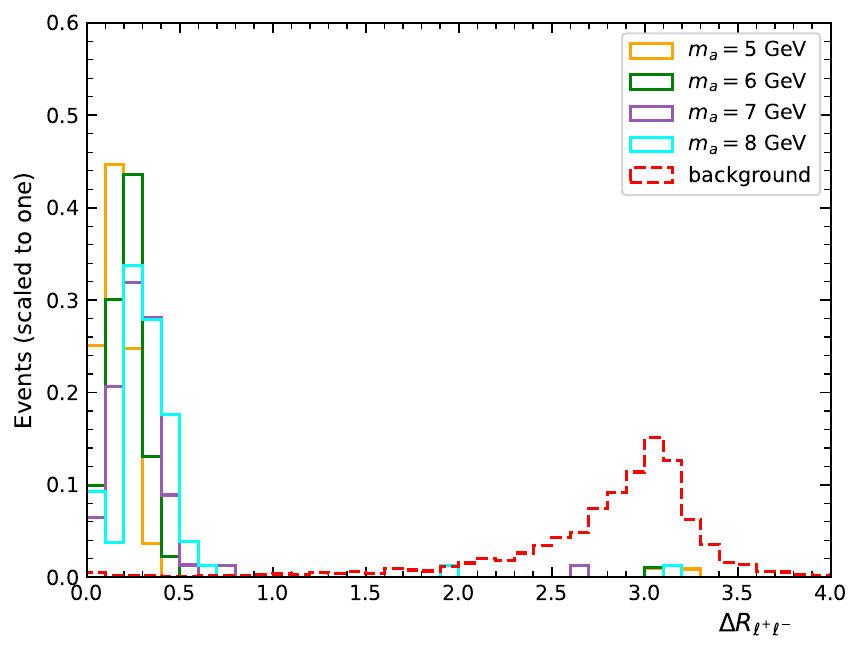}}
\subfigure[]{\includegraphics [scale=0.35] {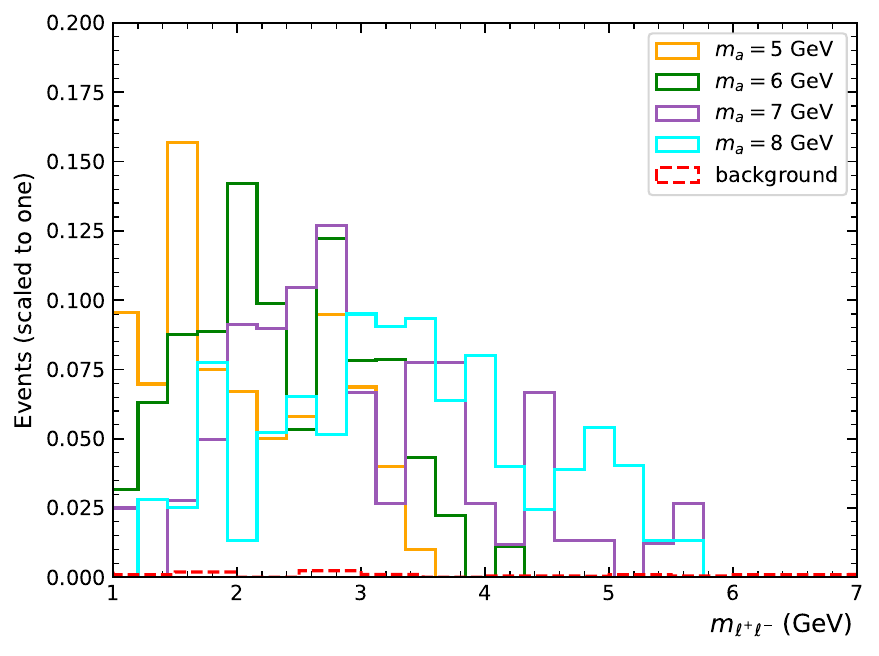}}
\caption{The normalized distributions of the observables $|d_0^{\ell}|$, $v_0^{\ell}$, $v_z^{\ell}$, $\Delta R_{\ell^+\ell^-}$ and $m_{\ell^+\ell^-}$ for the signal of selected ALP-mass benchmark points and the SM background for $g_{aWW}=5 \times 10^{-3}$ TeV$^{-1}$ at the 14 TeV HL-LHC with $\mathcal{L}=$ $3$ ab$^{-1}$.}
\label{fig:5}
\end{center}
\end{figure}

Unlike hadronic tau decays, which can be reconstructed as displaced $\tau$ jets, leptonic tau decays produce electrons or muons without forming visible tau jets. Consequently, the charged leptons in the final state serve as the primary observables. To retain a sufficient number of signal events, a pre-selection on the signal of lepton is also applied, requiring the transverse momentum $p_T^{\ell} > 10~\mathrm{GeV}$, the pseudorapidity $|\eta_{\ell}| < 2.4$ as well as the number of final-state leptons $N_{\ell^{+}} \geq 1$ and $N_{\ell^{-}} \geq 1$. To discriminate the signal from the SM background, we analyze several kinematic variables related to these charged leptons. These include the transverse impact parameter $d_0^\ell$, transverse distance $v_0^\ell$, longitudinal distance $v_z^\ell$, angular separation $\Delta R_{\ell^+\ell^-}$, and invariant mass $m_{\ell^+\ell^-}$. The normalized distributions of these variables at the 14 TeV HL-LHC with $\mathcal{L} = 3~\mathrm{ab}^{-1}$ are shown in Fig.~\ref{fig:5} for several benchmark ALP masses $m_a = 5$, 6, 7, and 8 GeV, where the solid and dashed lines respectively represent signal and SM background events. One can see that the signal events exhibit large values of $|d_0^\ell|$, $v_0^\ell$ and $v_z^\ell$, consistent with the long-lived nature of the ALP. In contrast, the charged leptons in the SM background are produced promptly at the interaction point, exhibiting minimal values for these displacement variables. The angular separation $\Delta R_{\ell^+\ell^-}$ effectively distinguishes the signal from the SM background. For the signal events, the two charged leptons from the decays of tau leptons produced by the light ALP are more collimated, resulting in smaller angular separations. In contrast, the SM background events tend to exhibit a wider separation. Furthermore, the invariant mass $m_{\ell^+\ell^-}$ serves as another effective kinematic variable to distinguish the signal from the SM background. For the signal events, the invariant mass distribution exhibits a clear peak slightly below the corresponding ALP mass, reflecting the energy carried away by neutrinos from the tau decays. With increasing ALP mass, the peak shifts to higher values and broadens gradually. By contrast, the SM background events lack such a distinct peak and instead display a more spread-out distribution.

Based on the features of the these kinematic distributions, a set of optimized cuts is applied to suppress the SM background and enhance the statistical significance of the signal, which are summarized in Table~\ref{tab3}.
To ensure that the ALP decays occur within the inner tracking detector, these parameters $|d_0^\ell|$, $v_0^\ell$, and $v_z^\ell$ are required to respectively satisfy the conditions $0.2 \, \text{mm} < |d_0^{\ell}| < 100 \, \text{mm}$, $v_0^{\ell} < 10 \, \text{cm}$ and $v_z^{\ell} < 30 \, \text{cm}$, which are also summarized in Table~\ref{tab3}. After applying the optimized cuts, the SM background is reduced to a level that can be considered negligible in the analysis. The cumulative selection efficiency $\epsilon$ and production cross sections of the $\ell^+ \ell^- \gamma \slashed{E}_T $ signal after imposing improved cuts for few ALP mass benchmark points and the coupling $g_{aWW}=5\times10^{-3}$ TeV$^{-1}$ at the $14$ TeV HL-LHC with $\mathcal{L}=$ $3$ ab$^{-1}$ are given in Table~\ref{tab4}. The values of $\epsilon$ can range from $4.68\times10^{-2}$ to $6.32\times10^{-2}$ for $m_a$ between 5 and 8 GeV. With effectively negligible background, the criterion $N_{sig} \geq 3$ corresponds to the 95$\%$ C.L. exclusion limit. We also apply the same search strategy given by the Eq.~\eqref{eq:11} over wider ranges of $m_a$ and $g_{aWW}$, deriving the expected 95$\%$ C.L. sensitivity of the 14 TeV HL-LHC with $\mathcal{L}=$ $3$ ab$^{-1}$ to long-lived ALPs in the blue region of the Fig.~\ref{fig:6}.

\begin{table}[H]
\begin{center}
\setlength{\tabcolsep}{1.9mm}{
\caption{Same as Table~\ref{tab1} but for the $\ell^+ \ell^- \gamma \slashed{E}_T$ signal.}
\label{tab3}
\resizebox{0.8\textwidth}{!}{
\begin{tabular}
[c]{c|c c c c c}\hline \hline
\multicolumn{1}{c|}{\textbf{Optimized cuts}} & \textbf{Selection criteria} \\
\hline
\makecell[l]{Pre-secetion: the transverse momentum, pseudorapidity and \\
                        multiplicity of the lepton $\ell~(\ell = e, \mu)$ in the final states} &
\makecell[c]{$p_T^{\ell} > 10~\mathrm{GeV}$, $|\eta_{\ell}| < 2.4$ \\
              $N_{\ell^{+}} \geq 1$, $N_{\ell^{-}} \geq 1$} \\
\makecell[l]{Cut 1: the transverse impact parameter} &
\makecell[c]{$0.2~\mathrm{mm} < |d_{\textnormal{0}}^{\ell}| < 100~\mathrm{mm}$} \\
\makecell[l]{Cut 2: the transverse and longitudinal distance} &
\makecell[c]{$v_{\textnormal{0}}^{\ell} < 10~\mathrm{cm}$, $v_z^{\ell} < 30~\mathrm{cm}$} \\
\makecell[l]{Cut 3: the angular separation between the two charged leptons} &
\makecell[c]{$\Delta R_{\ell^+\ell^-} < 1.0$} \\
\makecell[l]{Cut 4: the invariant-mass window cut on the charged lepton pair} &
\makecell[c]{$m_{\ell^+\ell^-} \leq 7~\mathrm{GeV}$} \\
\hline \hline
\end{tabular}}}
\end{center}
\end{table}

\begin{table}[H]\tiny
\begin{center}
\caption{Same as Table~\ref{tab2} but for the $\ell^+ \ell^- \gamma \slashed{E}_T$ signal.$~$\label{tab4}}
\newcolumntype{C}[1]{>{\centering\let\newline\\\arraybackslash\hspace{50pt}}m{#1}}
\begin{tabular}{m{2.0cm}<{\centering}|m{2.3cm}<{\centering} m{2.3cm}<{\centering} m{2.3cm}<{\centering}  m{2.3cm}<{\centering} m{2.3cm}<{\centering}}\hline \hline
\multirow{2}{*}{Selections} & \multicolumn{4}{c}{Cross sections~[pb] (efficiencies) for the signal} \\
\cline{2-5}
& $m_a=5$ GeV  & $m_a=6$ GeV  & $m_a=7$ GeV  & $m_a=8$ GeV  \\ \hline
Generator:  & \makecell{$2.41\times10^{-5}$} & \makecell{$2.40\times10^{-5}$} & \makecell{$2.40\times10^{-5}$} & \makecell{$2.39\times10^{-5}$} \\ \hline
Pre-secetion:  & \makecell{$2.41\times10^{-6}$\\($1.01\times10^{-1}$)} & \makecell{$2.37\times10^{-6}$\\($9.84\times10^{-2}$)} & \makecell{$2.02\times10^{-6}$\\($8.42\times10^{-2}$)} & \makecell{$2.11\times10^{-6}$\\($8.84\times10^{-2}$)} \\
Cut 1:       & \makecell{$1.95\times10^{-6}$\\($8.18\times10^{-2}$)} & \makecell{$1.86\times10^{-6}$\\($7.73\times10^{-2}$)} & \makecell{$1.58\times10^{-6}$\\($6.61\times10^{-2}$)} & \makecell{$1.66\times10^{-6}$\\($6.99\times10^{-2}$)} \\
Cut 2:       & \makecell{$1.12\times10^{-6}$\\($4.71\times10^{-2}$)} & \makecell{$1.25\times10^{-6}$\\($5.20\times10^{-2}$)} & \makecell{$1.29\times10^{-6}$\\($5.40\times10^{-2}$)} & \makecell{$1.51\times10^{-6}$\\($6.35\times10^{-2}$)} \\
Cut 3:       & \makecell{$1.11\times10^{-6}$\\($4.68\times10^{-2}$)} & \makecell{$1.25\times10^{-6}$\\($5.20\times10^{-2}$)} & \makecell{$1.29\times10^{-6}$\\($5.40\times10^{-2}$)} & \makecell{$1.51\times10^{-6}$\\($6.35\times10^{-2}$)} \\
Cut 4:       & \makecell{$1.11\times10^{-6}$\\($4.68\times10^{-2}$)} & \makecell{$1.25\times10^{-6}$\\($5.20\times10^{-2}$)} & \makecell{$1.28\times10^{-6}$\\($5.39\times10^{-2}$)} & \makecell{$1.50\times10^{-6}$\\($6.32\times10^{-2}$)} \\
\hline \hline
\end{tabular}
\end{center}
\end{table}

We plot the projected $95\%$ C.L. sensitivities of the 14 TeV HL-LHC with $\mathcal{L}=$ $3$ ab$^{-1}$ to long-lived ALPs in Fig.~\ref{fig:6}, where the orange and blue regions respectively represent the expected sensitivities from the $\pi^+ \pi^- \gamma \slashed{E}_T $ and $\ell^+ \ell^- \gamma \slashed{E}_T $ signals. For the $\pi^+ \pi^- \gamma \slashed{E}_T $ signal, the prospective sensitivities of the 14 TeV HL-LHC can reach $g_{aWW}\in [8.72 \times 10^{-3}, 6.42 \times 10^{-2}]$ TeV$^{-1}$ for the ALP mass $m_a \in [4, 10]$ GeV. While for the $\ell^+ \ell^- \gamma \slashed{E}_T $ signal, the 14 TeV HL-LHC can probe a broader parameter space, with the projected sensitivities covering $g_{aWW} \in [4.17 \times 10^{-3}, 2.00 \times 10^{-1}]$ TeV$^{-1}$ for the same ALP mass range. These results demonstrate that long-lived ALPs can be effectively probed via the $\pi^+ \pi^- \gamma \slashed{E}_T $ and $\ell^+ \ell^- \gamma \slashed{E}_T $ signals at the 14 TeV HL-LHC with $\mathcal{L}=$ $3$ ab$^{-1}$.

\begin{figure}[H]
\begin{center}
\centering\includegraphics [scale=0.35] {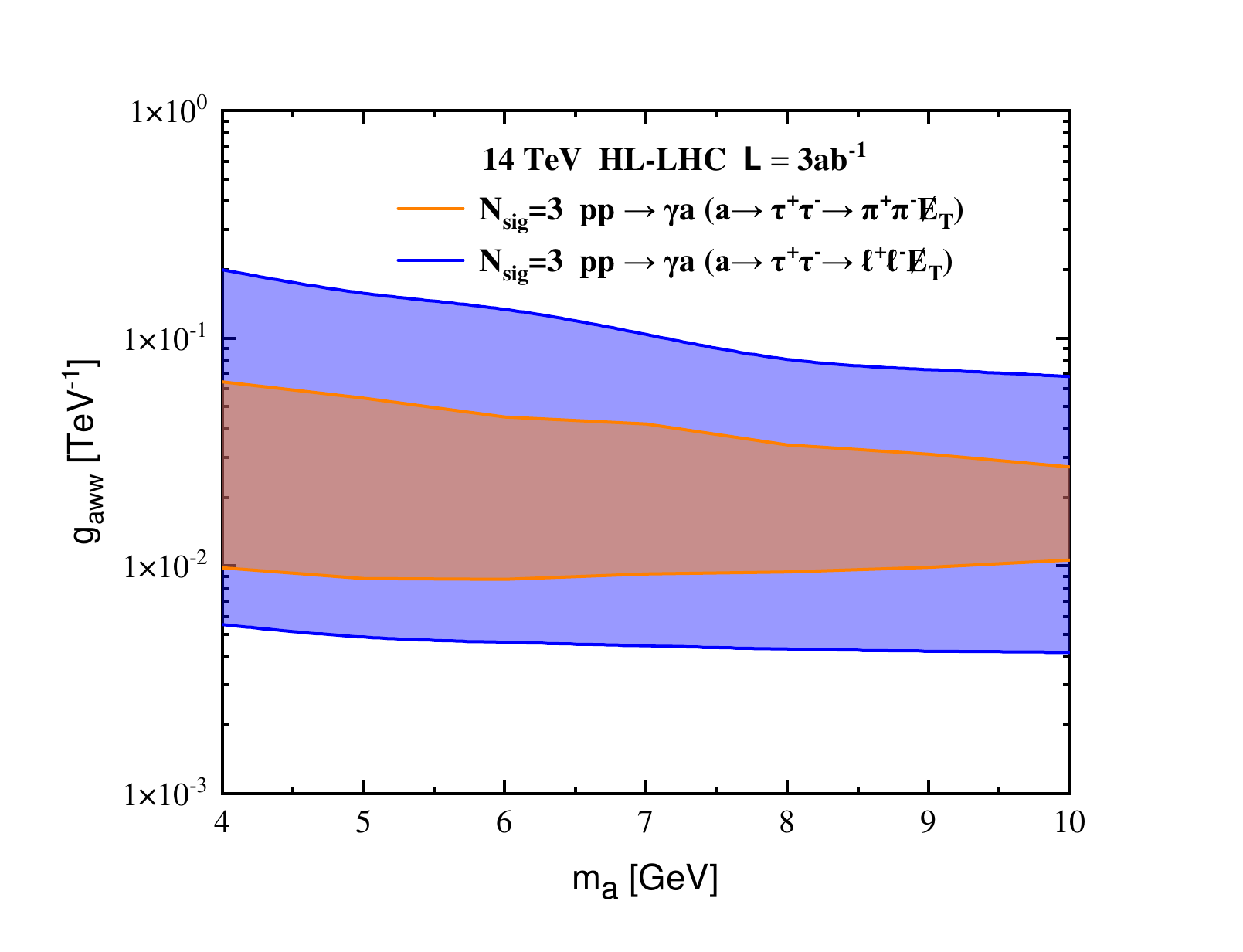}
\caption{The projected $95\%$ C.L. sensitivities of the 14 TeV HL-LHC with $\mathcal{L}=$ $3$ ab$^{-1}$ to long-lived ALPs via the $\pi^+ \pi^- \gamma\slashed{E}_T $ signal~(orange region) and the $\ell^+ \ell^- \gamma \slashed{E}_T $ signal~(blue region).}
\label{fig:6}
\end{center}
\end{figure}

\begin{figure}[H]
\begin{center}
\centering\includegraphics [scale=0.4] {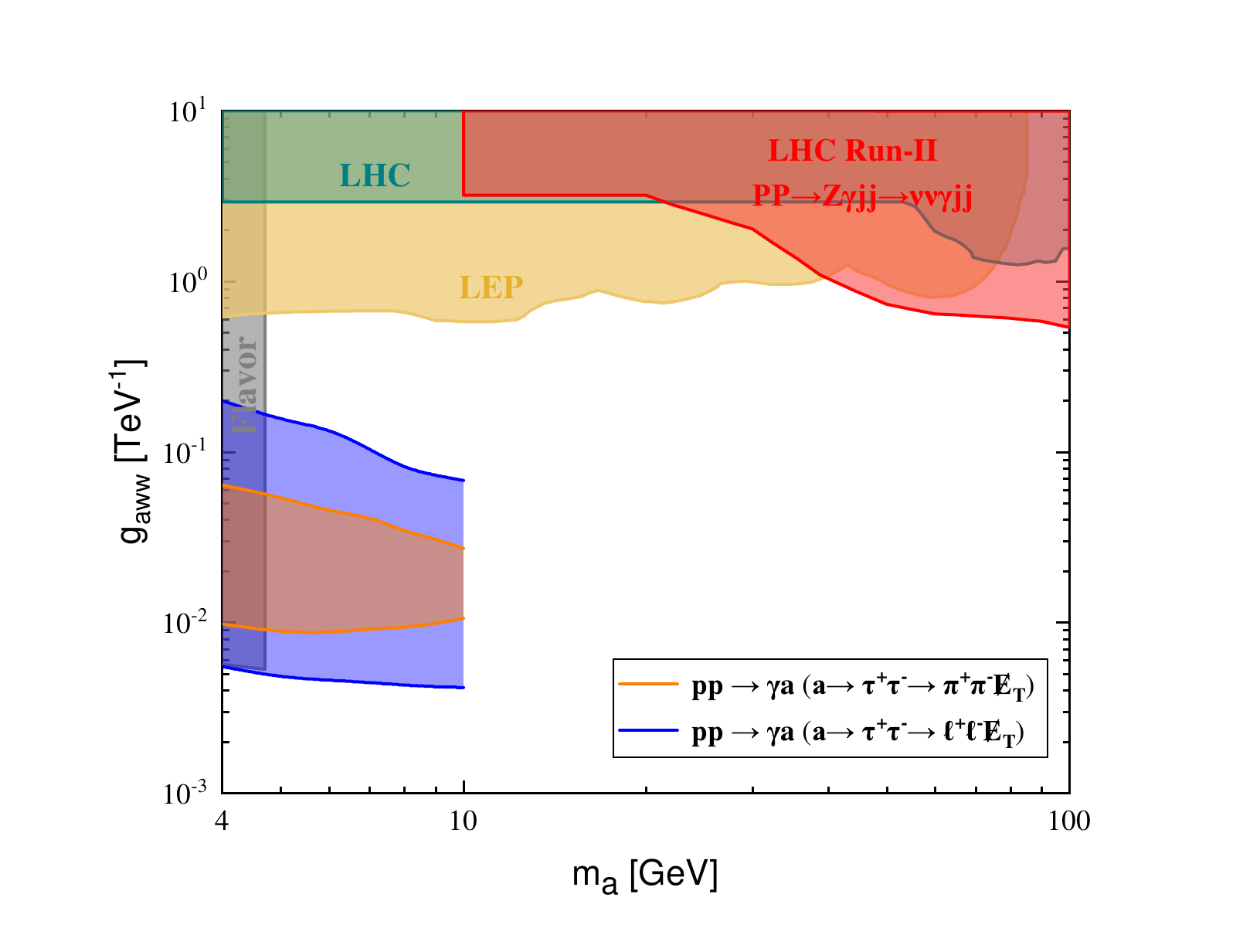}
\caption{Our projected $95\%$ C.L. sensitivities of the 14 TeV HL-LHC with $\mathcal{L}=$ $3$ ab$^{-1}$ to long-lived ALPs via the $\pi^+ \pi^- \gamma \slashed{E}_T $ signal~(orange region) and the $\ell^+ \ell^- \gamma \slashed{E}_T $ signal~(blue region) and other current exclusion regions.}
\label{fig:7}
\end{center}
\end{figure}

The projected $95\%$ C.L. sensitivities of the $14$ TeV HL-LHC with $\mathcal{L}=$ $3$ ab$^{-1}$ to long-lived ALPs for the $\pi^+ \pi^- \gamma \slashed{E}_T $ signal~(orange region) and the $\ell^+ \ell^- \gamma \slashed{E}_T $ signal~(blue region) derived in this paper and other exclusion regions from previous studies are given in Fig.~\ref{fig:7}. The gray region labeled ``Flavor" represents the constraints obtained by the LHCb collaboration via rare $B$ meson decay processes $B^{\pm} \to K^{\pm}a$~\cite{LHCb:2016awg} and $B^{0} \to K^{0*}a$~\cite{LHCb:2015nkv} with $a\to \mu^+ \mu^-$ for $0.25$ GeV $\leq m_a \leq 4.7$ GeV and $0.21$ GeV $\leq m_a \leq 4.35$ GeV. The yellow region labeled ``LEP" indicates the limits from exotic $Z$-boson decays at the LEP for $4.8 \, \text{GeV} \leq m_a \leq m_Z$, including the constraints from the OPAL search via the process $Z \to \gamma a \to \gamma\gamma + \slashed E_T$~\cite{OPAL:1993ezs} and the L3 search via the process $Z \to \gamma a \to \gamma jj$~\cite{L3:1992kcg}. These LEP searches are sensitive to promptly decaying ALPs and therefore probe ALP parameter space with relatively large couplings. In contrast, our work focused on long-lived ALPs characterized by significantly weaker couplings, evading these prompt decay constraints. To quantitatively substantiate this, we compute the branching fraction of the exotic decay $Z \to \gamma a$. The decay width is given by $\Gamma(Z\to a \gamma) =  \frac{m_Z^3 s_{W}^2}{96\pi c_{W}^2}|g_{aWW}|^2 (1-\frac{m_a^2}{m_Z^2})^3$. For a representative coupling $g_{aWW} \sim 1 \times 10^{-2}~\text{TeV}^{-1}$ in the sensitivity region of our analysis, the branching fraction $\Gamma(Z \to \gamma a) / \Gamma_Z$ is approximately $3 \times 10^{-8}$, which is well below the upper limits given by the OPAL and L3 collaborations, $\mathrm{Br}(Z \to \gamma a \to \gamma\gamma + \text{invisible}) < 3.1 \times 10^{-6}$~\cite{OPAL:1993ezs} and $\mathrm{Br}(Z \to \gamma a \to \gamma jj) < 1 \times 10^{-4}$~\cite{L3:1992kcg}. Therefore, the parameter space probed by our displaced vertex analysis at the HL-LHC remains unconstrained by exotic $Z$-boson decay searches. The green region labeled ``LHC" represents the constraints from various LHC searches~\cite{Craig:2018kne}. Additionally, the red region indicates the limits from 13 TeV Run-II searches via the process $pp \to a jj \to Z(\to \nu \bar{\nu}) \gamma jj$ at the LHC~\cite{Aiko:2024xiv}.

Comparing the prospective sensitivities of the 14 TeV HL-LHC with exclusion regions from other experiments, we find that the sensitivities obtained through the $\pi^+ \pi^- \gamma \slashed{E}_T $ and $\ell^+ \ell^- \gamma \slashed{E}_T $ signals can probe the regions with smaller ALP masses and weaker couplings. In contrast, prompt decay analyses from LEP and LHC experiments are limited to probing the regions with larger ALP masses or stronger couplings. Consequently, our analysis complements previous studies, providing access to the regions with smaller couplings that are inaccessible to prompt decay searches. Furthermore, compared to the exclusion limits on long-lived ALPs obtained by the LHCb collaboration through rare $B$ meson decays, our results can probe a broader parameter space. Therefore, we conclude that the 14 TeV HL-LHC with $\mathcal{L}=$ $3$ ab$^{-1}$ has significant potential for detecting long-lived ALPs via their displaced vertex signatures.

%%%%%%%%%%%%%%%%%%%%%%%%%%%%%%%%%%%%%%%%%%%%%%%%%%%%%%%%%%%%%%%%%%%%%%%%%%%%%%%%%%%%%%%%%%%%%%%%%%%%%%%%%%%%%%%%%%%%%%%%%%%%%%%%%%%%
\section{Conclusions and Discussions}

Axion-like particles (ALPs) are well-motivated low-energy relics arising from high-energy extensions of the standard model (SM). The masses of ALPs and their couplings to the SM particles are considered to be independent parameters, thus they would generate rich and interesting phenomenology in current and future experiments. Many searches focus on promptly decaying ALPs, which are typically associated with larger masses or stronger couplings. However, as the ALP mass and coupling decrease, its proper decay length $c\tau$ increases, making ALPs long-lived and evading prompt decay searches. In this paper, we explore the parameter space where the ALP is sufficiently long-lived to evade prompt decay searches but decays within the inner tracking detector, producing displaced vertices associated with its charged decay products. The HL-LHC, with high integrated luminosity and upgraded tracking systems, provides a promising environment to probe such long-lived ALPs.

In this work, we explore the possibility of detecting long-lived ALPs in the photophobic scenario for the ALPs with masses in the range of $4-10$ GeV within the inner tracking detector at the 14 TeV HL-LHC with $\mathcal{L} = 3~\mathrm{ab}^{-1}$. We consider the process $pp \to \gamma a$ with $a$ decaying into a pair of displaced charged leptons. We find that the expected sensitivity to long-lived ALPs via displaced muon signatures has been covered by the LHCb searches. Therefore, we turn our attention to exploring long-lived ALPs via displaced tau signatures and simulate their subsequent hadronic and leptonic decays. We have performed Monte Carlo simulations for the $\pi^+ \pi^- \gamma \slashed{E}_T $ and $\ell^+ \ell^- \gamma \slashed{E}_T $ signals and obtained prospective $95\%$ C.L. sensitivities of the HL-LHC to long-lived ALPs. For the $\pi^+ \pi^- \gamma \slashed{E}_T $ signal, the HL-LHC can probe values of $g_{aWW}$ from $8.72 \times 10^{-3}$ TeV$^{-1}$ to $6.42 \times 10^{-2}$ TeV$^{-1}$ for ALP with masses between 4 and 10 GeV. The HL-LHC can explore a broader parameter space via the $\ell^+ \ell^- \gamma \slashed{E}_T $ signal, with sensitivities extending to values of $g_{aWW}$ from $4.17 \times 10^{-3}$ TeV$^{-1}$ to $2.00 \times 10^{-1}$ TeV$^{-1}$.
Comparing our numerical results with exclusion regions from other experiments, we find that the HL-LHC can probe regions of the ALP parameter space with smaller masses and weaker couplings that are not accessible through some prompt decay analyses from LEP and LHC experiments. Furthermore, compared to the exclusion limits obtained by the LHCb collaboration via rare $B$ meson decays, our results can probe a broader parameter space. Thus, we can conclude that the HL-LHC has a great potential to detect the long-lived ALPs in the photophobic scenario via their displaced vertex signatures.

In this paper, we primarily explore the potential of the HL-LHC for detecting long-lived particles.
Lepton colliders can also offer promising prospects for detecting long-lived ALPs due to their cleaner experimental environment with reduced QCD background. However, their center-of-mass energy is typically lower than that of the HL-LHC. Additionally, the LHCb can also provide valuable complementary coverage in the ALP mass range considered here, benefiting from excellent tau and pion reconstruction capabilities as well as lower trigger thresholds~(see, e.g., Refs.~\cite{CidVidal:2018blh,BuarqueFranzosi:2021kky}). However, its integrated luminosity is considerably lower than that of the HL-LHC, which may limit its sensitivity to the signal processes considered in this paper. We will investigate the feasibility of detecting long-lived ALPs at the lepton colliders and LHCb
in future studies.

%%%%%%%%%%%%%%%%%%%%%%%%%%%%%%%%%%%%%%%%%%%%%%%%%%%%%%%%%%%%%%%%%%%%%%%%%%%%%%%%%%%%%%%%%%%%%%%%%%%%%%%%%%%%%%%%%%%%%%%%%%%%%%%%%%%%
\section*{ACKNOWLEDGMENT}

This work was partially supported by the National Natural Science Foundation of China under Grants No. 11875157, No. 12147214 and No. 12575106.

%%%%%%%%%%%%%%%%%%%%%%%%%%%%%%%%%%%%%%%%%%%%%%%%%%%%%%%%%%%%%%%

\end{document}